\documentclass
[a4paper]{llncs}
\usepackage{graphicx}
\usepackage[utf8]{inputenc}
\usepackage{amssymb,amsmath,array,url}

\usepackage{subfig}
\usepackage{microtype}

\begin{document}
\mainmatter
\title{Visual Mining of Epidemic Networks}
\author{St\'ephan Cl\'emen\c{c}on\inst{1}\thanks{This work was supported by
    the French Agency for Research under grant ANR Viroscopy
    (ANR-08-SYSC-016-03) and by AECID project D/030223/10} 
\and Hector De Arazoza\inst{2,3} \and
  Fabrice Rossi\inst{1} \and Viet-Chi Tran\inst{3}}
\institute{Institut T\'el\'ecom, T\'el\'ecom ParisTech, LTCI - UMR CNRS 5141
\\46, rue Barrault, 75013 Paris -- France
\and Facultad de Matem\'atica y Computaci\'on, Universidad de la Habana,
\\  La Habana, Cuba
\and 
Laboratoire Paul Painlev\'e UMR CNRS No. 8524, Universit\'e Lille
  1, \\59 655 Villeneuve d'Ascq Cedex, France
\\
\email{stephan.clemencon@telecom-paristech.fr, arazoza@matcom.uh.cu,
  fabrice.rossi@telecom-paristech.fr, chi.tran@math.univ-lille1.fr}
}
\maketitle

\begin{abstract}
  We show how an interactive graph visualization method based on maximal
  modularity clustering can be used to explore a large epidemic network. The
  visual representation is used to display statistical tests results that
  expose the relations between the propagation of HIV in a sexual contact
  network and the sexual orientation of the patients.
\end{abstract}

\section{Introduction}
Large graphs and networks are natural mathematical models of interacting
objects such as computers on the Internet or articles in citation
networks. Numerous examples can be found in the biomedical context from
metabolic pathways and gene regulatory networks to neural networks
\cite{Newman2003GraphSurveySIAM}. The present work is dedicated to one type of
such biomedical network, namely epidemic networks \cite{KeelingEames2005}:
such a network models the transmission of a directly transmitted infectious
disease by recording individuals and their contacts, other individuals to whom
they can pass infection.

Understanding the dynamic of the transmission of diseases on real world
networks can lead to major improvements in public health by enabling effective
disease control thanks to better information about risky behavior, targeted
vaccination campaigns, etc. While transmissions can be studied on artificial
networks, e.g., some specific types of random networks
\cite{KeelingEames2005}, such networks fail to exhibit all the characteristics
observed in real social networks (see
e.g.\cite{Newman2003GraphSurveySIAM}). It is therefore important to get access
to and to analyze large and complex real world epidemic networks. As pointed
out in \cite{KeelingEames2005}, the actual definition of the social network on
which the propagation takes place is difficult, especially for airborne
pathogens, as the probability of disease transmission depends strongly on the
type of interaction between persons. This explains partially why sexually
transmitted diseases (STD) epidemic networks have been studied more frequently
than other networks \cite{KeelingEames2005,LiljerosEtAl2003SexualNetworks}.

We study in this paper a large HIV epidemic network that has some unique
characteristics: it records almost 5400 HIV/AIDS cases in Cuba from 1986 to
2004; roughly 2400 persons fall into a single connected component of the
infection network. STD networks studied in the literature are generally
smaller and/or do not exhibit such a large connected component and/or contain
a very small number of infected persons. For instance, the Manitoba study (in
Canada, \cite{WylieJolly2001}) covers 4544 individuals with some STD, but the
largest connected component covers only 82 persons. The older Colorado Springs
study \cite{RothenbergEtAl1995} covers around 2200 persons among which 965
falls in connected component (the full network is larger but mixes sexual
contacts and social ones; additionally, the sexual networks contains only a
very small number of HIV positive persons).

While the large size and coverage of the studied network is promising, it has
also a main negative consequence: manual analysis and direct visual
exploration, as done in e.g. \cite{LiljerosEtAl2003SexualNetworks}, is not
possible. We propose therefore to analyze the network with state-of-the-art
graph visualization methods \cite{ClemenconEtAlESANN2011}.

We first describe the epidemic network in Section \ref{sec:cuban-hiva-datab}
and give an example of the limited possibilities of macroscopic analysis on
this dataset. Then Section \ref{sec:visual-mining} recalls briefly the visual
mining technique introduced in \cite{ClemenconEtAlESANN2011} and shows how it
leads to the discovery of two non obvious sub-networks with distinctive
features. 

\section{Cuban HIV/AIDS Database}\label{sec:cuban-hiva-datab}
The present work studies an anonymized national dataset which lists 5389
Cuban residents with HIV/AIDS, detected between 1986 and 2004. Each patient is
described by several variables including gender, sexual orientation, age at
HIV/AIDS detection, etc. (see \cite{Auvert} for details.) 

\subsection{Data collection}
The Cuban HIV/AIDS program produces this global monitoring using several
sources that range from systematic testing of pregnant women and all blood
donations to general practitioner testing recommendations. In
addition, the program conducts an extended form of infection tracing that
leads to the epidemic network studied in this work.

Indeed, each new infected patient is interviewed by health workers and invited
to list his/her sexual partners from the last two years. The primary use of
this approach is to discover potentially infected persons and to offer them
HIV testing. An indirect result is the construction of a network of infected
patients. Sexual partnerships are indeed recorded in the database for all
infected persons. Additionally, a probable infection date and a transmission
direction are inferred from other medical information, leading
to a partially oriented infection network. While this methodology is not
contact tracing \emph{stricto sensu} as non infected patients are not included
in the database (contrarily to e.g. \cite{RothenbergEtAl1995}), the program
records the total number of sexual partners declared for the two years period
as well as a few other details, leading to an extended form of infection
tracing. (see \cite{KeelingEames2005} for differences between contact and
infection tracing.)

\subsection{Macroscopic analysis}
The 5389 patients are linked by 4073 declared sexual relations among which
2287 are oriented by transmission direction. A significant fraction of the
patients (44 \%) belong to a giant connected component with 2386 members. The
rest of the patients are either isolated (1627 cases) or members of very small
components (the second largest connected component contains only 17
members). 

As the sexual behavior has a strong influence on HIV transmission, it seems
important to study the relations between the network structure and sexual
orientation of the patients. In the database, female HIV/AIDS patients are all
considered to be heterosexual as almost no HIV transmission between female has
been confirmed \cite{KwakwaGhobrial2003}. Male patients are
categorized into heterosexual man and ``Man having sex with Men'' (MSM); the
latter being men with at least one male sexual partner identified during their
interview. 

\begin{figure}[htbp]
  \centering
\includegraphics[width=0.9\linewidth]{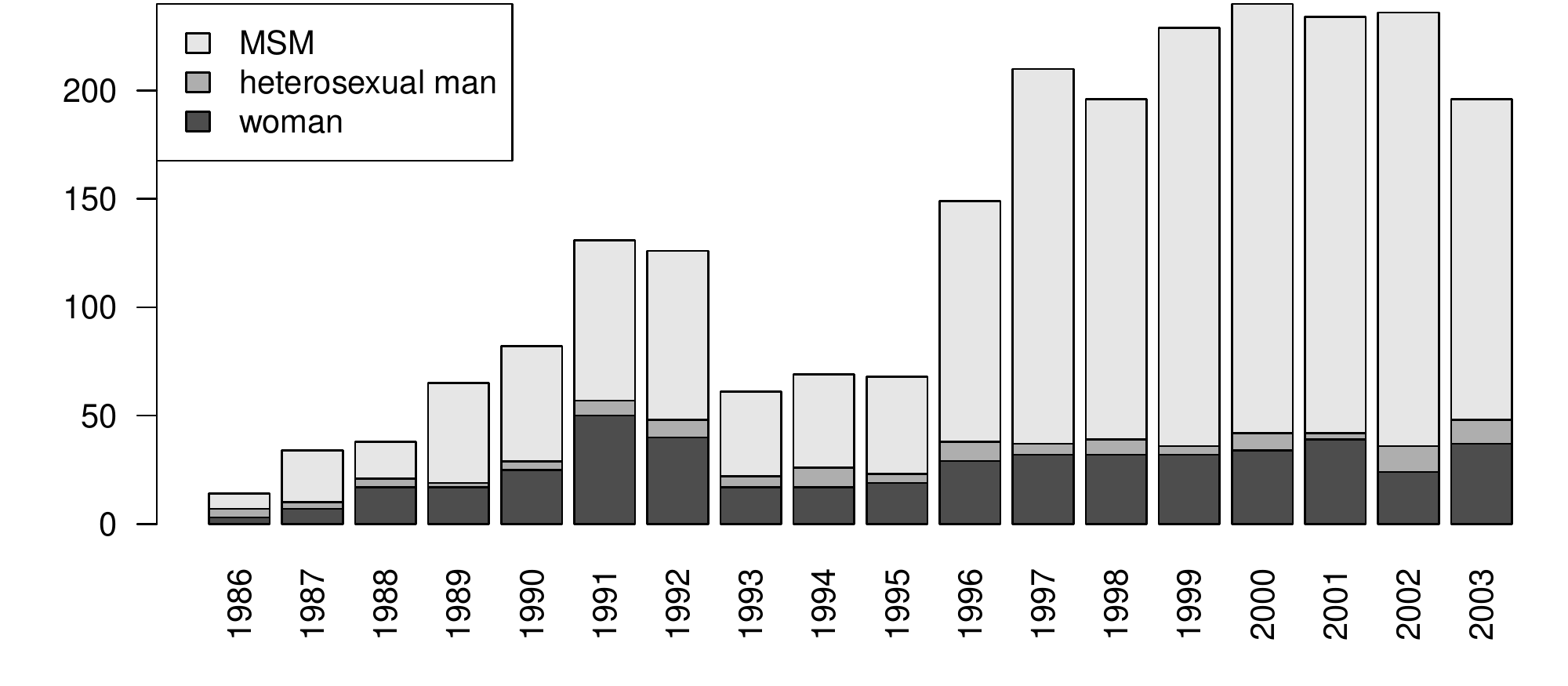}
  \caption{Yearly sexual orientation distribution in the giant component}
  \label{fig:so:year}
\end{figure}

The distributions of genders and of sexual orientations is given in Table
\ref{tab:Gender:SO}: the giant component contains proportionally more MSM than
the full population; this seems logical because of the higher probability of
HIV transmission between men \cite{VargheseEtAl2002}. 
\begin{table}[htbp]
  \centering
\begin{tabular}{rrrrr}
  \hline
& \multicolumn{2}{c}{\textbf{full network}}&\multicolumn{2}{c}{\textbf{giant component}}\\
 & absolute & relative &absolute & relative\\ 
  \hline
woman & 1109 & 0.21 & 472 & 0.20\\ 
  heterosexual man & 566 & 0.11 & 110 & 0.05\\ 
  MSM & 3714 & 0.69& 1804 & 0.76  \\ 
   \hline
\end{tabular}    
  \caption{Gender and sexual orientation distributions in the whole network and
  in the giant component}
  \label{tab:Gender:SO}
\vskip -1.2em
\end{table}
Additionally, as shown on Figure \ref{fig:so:year}, the sexual orientation
of the newly detected patients changes through years (note that year 2004 is
incomplete and therefore has not been included in this analysis): the
percentage of heterosexual men seems to decrease. This is also 
the case for women, but to a lesser extent. 

However, the evolution of the percentage of MSM through time is difficult to
analyze at a macroscopic level. For instance, multiple explanations can be
offered to explain the influence of the sexual orientation of the patients on
their average shortest path distances in the graph (see Table
\ref{tab:geo:so}).
\begin{table}[htbp]
  \centering
\begin{tabular}{rccc}
  \hline
 & MSM & heterosexual man & woman \\ 
  \hline
MSM & 10.30 & 10.83 & 10.24 \\ 
  heterosexual man & 10.83 & 9.87 & 9.30 \\ 
  woman & 10.24 & 9.30 & 8.76 \\ 
   \hline
 \end{tabular}
 \caption{Average shortest path distances in the giant component of the infection network, conditioned by
   the sexual orientation of the extremal points (the global average is
   10.24)}\label{tab:geo:so}
\vskip -1.2em
\end{table}
One the one hand, temporal distance should be reflected by distance in the
graph as the infection tracing has a short time span. Then one might expect
short paths between heterosexual men as they should be concentrated in the
early phase of the epidemic and long chains should be rare. However, on the
other hand, heterosexual men could also be linked to the network mostly
through women. Then their relative short distances could be explained via the
relative short distances between women themselves and the time aspect would
play no role in the observed distances. 

\section{Visual Mining}\label{sec:visual-mining}
Because of the size of the giant component, a direct visual analysis is
impossible and the interplay between sexual orientation and infection is
difficult to analyze.  We show in this section how the difficulty can be
circumvented by combining \emph{clustered graph} visualization techniques
\cite{DBLP:conf/gd/EadesF96} with efficient maximal modularity graph
clustering \cite{NoackRotta2009MultiLevelModularity}, as proposed in
\cite{ClemenconEtAlESANN2011}. The methodology introduced here can be used to
analyze the relation between other quantities and the infection structure.

\subsection{Methodology}
This section briefly explains the graph visualization method used to analyze
the infection network. Details can be found in
\cite{ClemenconEtAlESANN2011}. 

\begin{figure}
  \centering
\subfloat[Best clustering]{\includegraphics[width=0.45\linewidth]{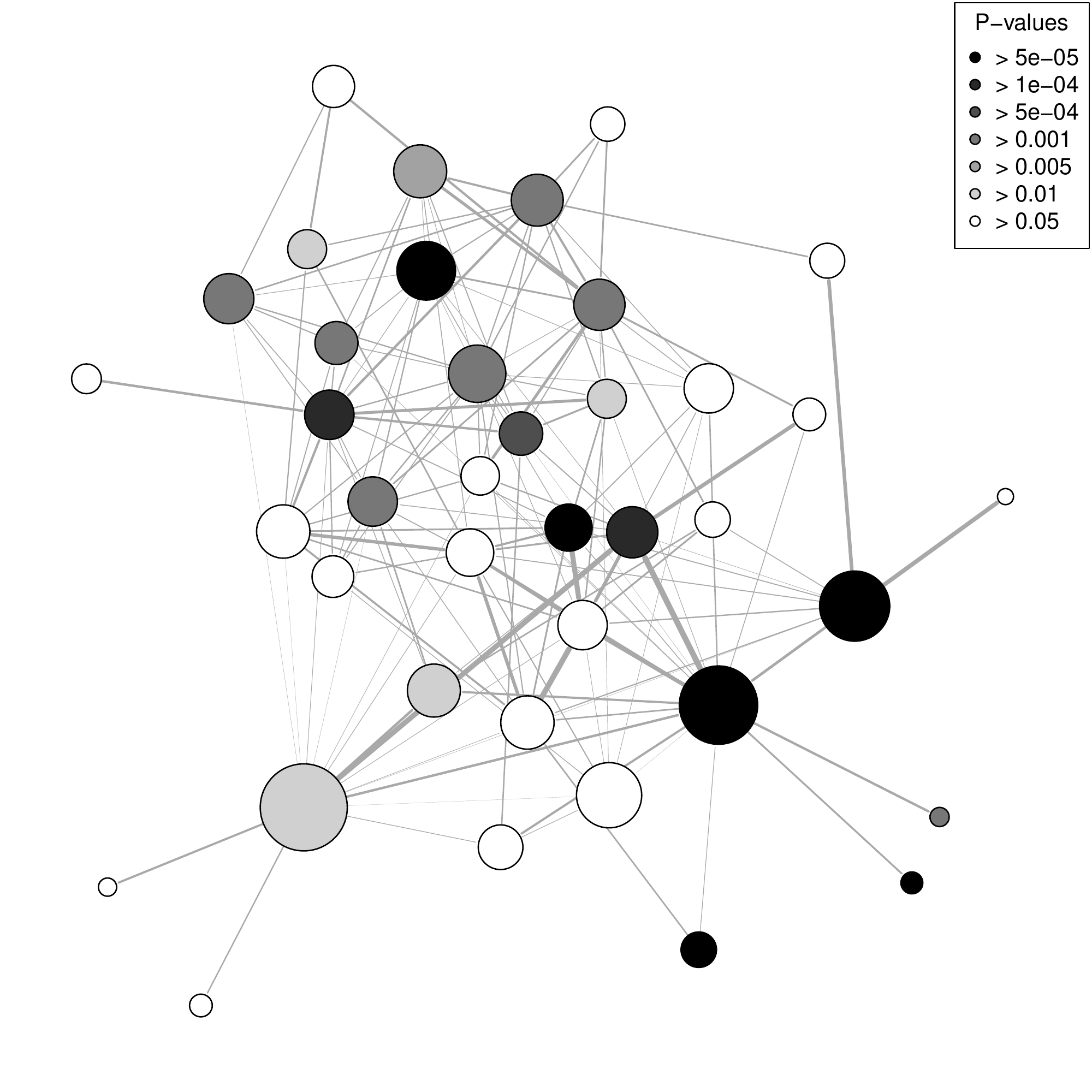}}%
\quad \subfloat[Maximally refined clustering]{\includegraphics[width=0.45\linewidth]{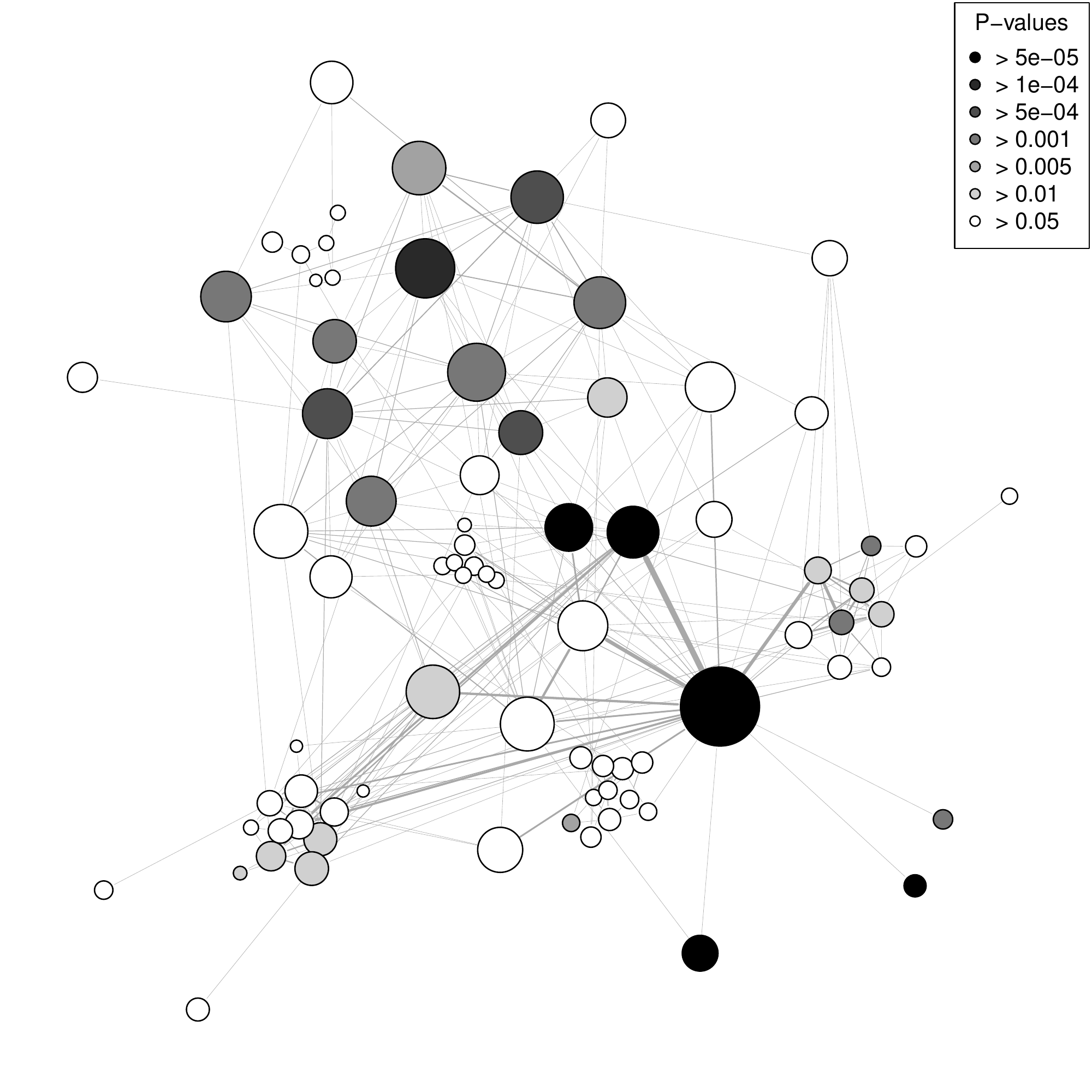}}  
\caption{Clustered graph visualization}
  \label{fig:clustered:graph}
\end{figure}

The classical strategy used to display a large network (e.g., with more than a
hundred nodes) is to coarsen the network via a clustering method, leading to
the so-called \emph{clustered graph} visualization problem
\cite{DBLP:conf/gd/EadesF96}. To implement this strategy, we use a maximal
modularity graph clustering approach
\cite{NoackRotta2009MultiLevelModularity}, as maximizing the modularity leads
in general to meaningful clusters \cite{FortunatoSurveyGraphs2010} which are
additionally well adapted to visualization \cite{Noack2009}. Then rather than
displaying the original network, we use the standard Fruchterman Reingold
algorithm \cite{FruchtermanReingoldGraph1991} to display the network of
clusters. Figure \ref{fig:clustered:graph} gives concrete examples of the
results: in the present context, each cluster consists of a group of patients
linked by sexual relationships (the size of the group is represented by the
surface of the disk used in the figures). Two groups are linked in the display
when there is at least one sexual partnership between patients of the two
groups. The thickness of the link encodes the number of between group sexual
contacts. 

We implemented the hierarchical principle used in \cite{DBLP:conf/gd/EadesF96}
by providing interactive coarsening and refining of the clustering. Indeed the
best clustering of the network might be too coarse to give insights on the
structure of network or too fine to lead to a legible drawing. Coarsening is
implemented by a greedy merging of clusters (as is used in
\cite{NoackRotta2009MultiLevelModularity}) while refinement is obtained by
applying maximal modularity clustering to each sub-cluster, taken in isolation
from the rest of the network. We keep only statistically significant
coarsenings and refinements: the modularity of the selected clusterings must be
higher than the maximal modularity obtained on random graphs with the same
degree distribution (see \cite{ClemenconEtAlESANN2011} for details). Figure
\ref{fig:clustered:graph} (b) gives an example of a refinement for the
clustering used in Figure \ref{fig:clustered:graph} (a), while Figure
\ref{fig:coarsened:graph} is based on a coarsening of the clustering. 

\subsection{Results}
Using \cite{NoackRotta2009MultiLevelModularity}, we obtain a partition of the
giant component into 39 clusters, with a modularity of 0.85. This is
significantly higher than the modularities of random graphs with identical
sizes and degree distributions: the highest value among 50 random graphs is
0.74. The corresponding layout is given by Figure \ref{fig:clustered:graph}
(a). We use this layout as a support for visualization exploration of the
sexual orientation distribution: nodes are darkened according to the $p$ value
of a chi squared test conducted on the distribution of the sexual orientation
of persons in each cluster versus the distribution of the same variable in the
full connected component. It appears clearly that some clusters have a specific
distribution of the sexual orientation variable.

The possibilities for refining the clustering in this case are quite limited:
only 5 of the 39 clusters have a significant substructure. Nevertheless,
Figure \ref{fig:clustered:graph} (b), which shows the fully refined graph
(with modularity 0.81) gives interesting insights on the underlying graph. For
instance, an upper left gray cluster is split into 6 white clusters: while the
best clustering of those persons leads to an atypical sexual orientation
distribution, this is not the case of each sub-cluster. This directs the
analyst to a detailed study of the corresponding persons: it turns out that
the cluster consists mainly in MSM patients. Sub-clusters are small enough
($\sim$ 7 patients) for MSM dominance to be possible by pure chance, while
this is far less likely for the global cluster with 41 patients (among which
39 are MSM).

\begin{figure}
  \centering
\subfloat[Chi square P values]{\includegraphics[width=0.45\linewidth]{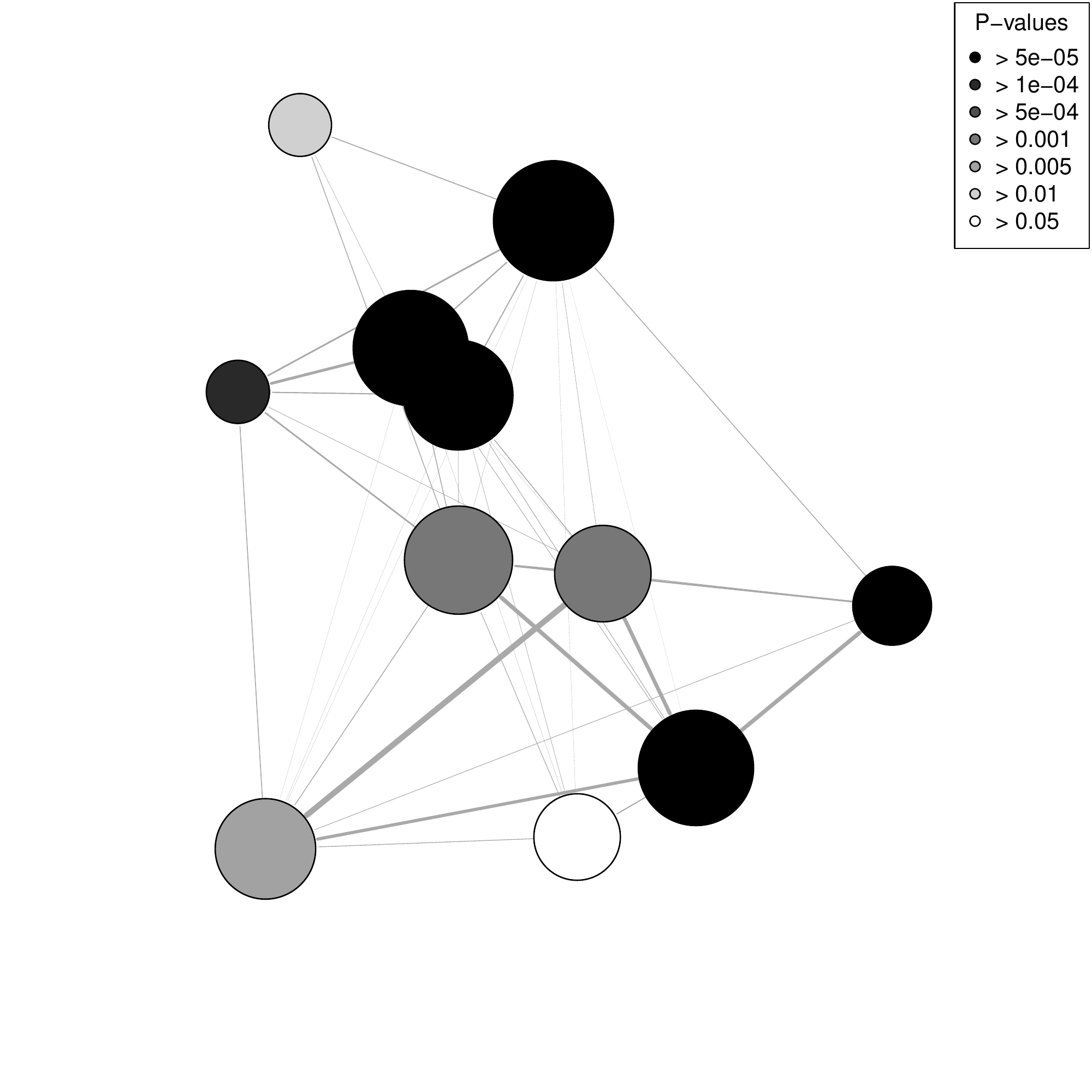}}%
\quad \subfloat[Pearson's residuals for MSM]{\includegraphics[width=0.45\linewidth]{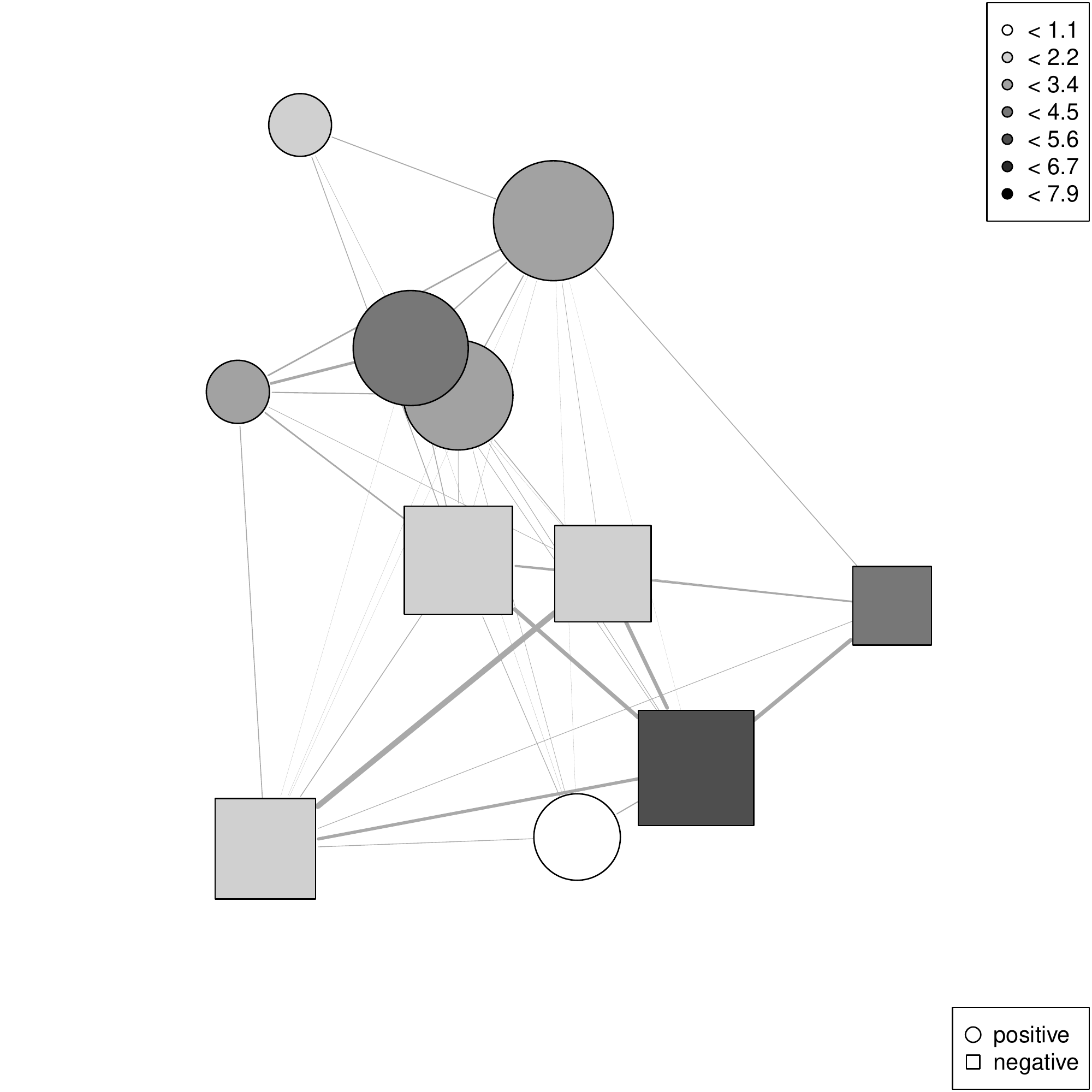}}  
\caption{Coarsened clustered graph visualization}
  \label{fig:coarsened:graph}
\end{figure}

Coarsening can be done more aggressively on this graph: clusterings down to 8
clusters have modularity above the random level. With 11 clusters, the
modularity reaches 0.81, a similar value as the maximally refined graph. While
Figure \ref{fig:clustered:graph} (a) is legible enough to allow direct
analysis, the coarsening emphasizes the separation of the graph into two
sparsely connected structures with mostly atypical sexual orientation
distributions in the associated clusters, as shown in Figure
\ref{fig:coarsened:graph} (a). Figure \ref{fig:coarsened:graph} (b) represents
the Pearson's residuals of the chi square tests for the MSM
sexual orientation: it clearly shows that a part of the largest connected
component contains more than expected MSM (circle nodes) while the
other part contains less than expected (square nodes). 

This finding directs the analyst to a sub-population study. The original 39
clusters are merged into three groups: MSM clusters (atypical clusters in the
upper part of the graph which contain more MSM than expected), Mixed clusters
(atypical clusters in the lower part of the graph, which contain less MSM
than expected) and typical clusters. Then the geodesic analysis summarized in
Table \ref{tab:geo:so} is done at this group level, leading to Table
\ref{tab:geo:groups}.
\begin{table}[htbp]
\centering
\begin{tabular}{rccc}
  \hline
 & MSM clusters & Mixed clusters & Typical clusters\\ 
  \hline
MSM clusters & 9.79 & 12.28 & 11.93 \\ 
  Mixed  clusters& 12.28 & 7.56 & 9.24 \\ 
  Typical  clusters& 11.93 & 9.24 & 12.04 \\ 
   \hline
\end{tabular}
  \caption{Average geodesic distances between members of the three cluster groups. Paths are restricted to patients belonging to the groups under consideration}
  \label{tab:geo:groups}
\vskip -1.2em
\end{table}

\begin{figure}[htbp]
  \centering
\includegraphics[width=0.9\linewidth]{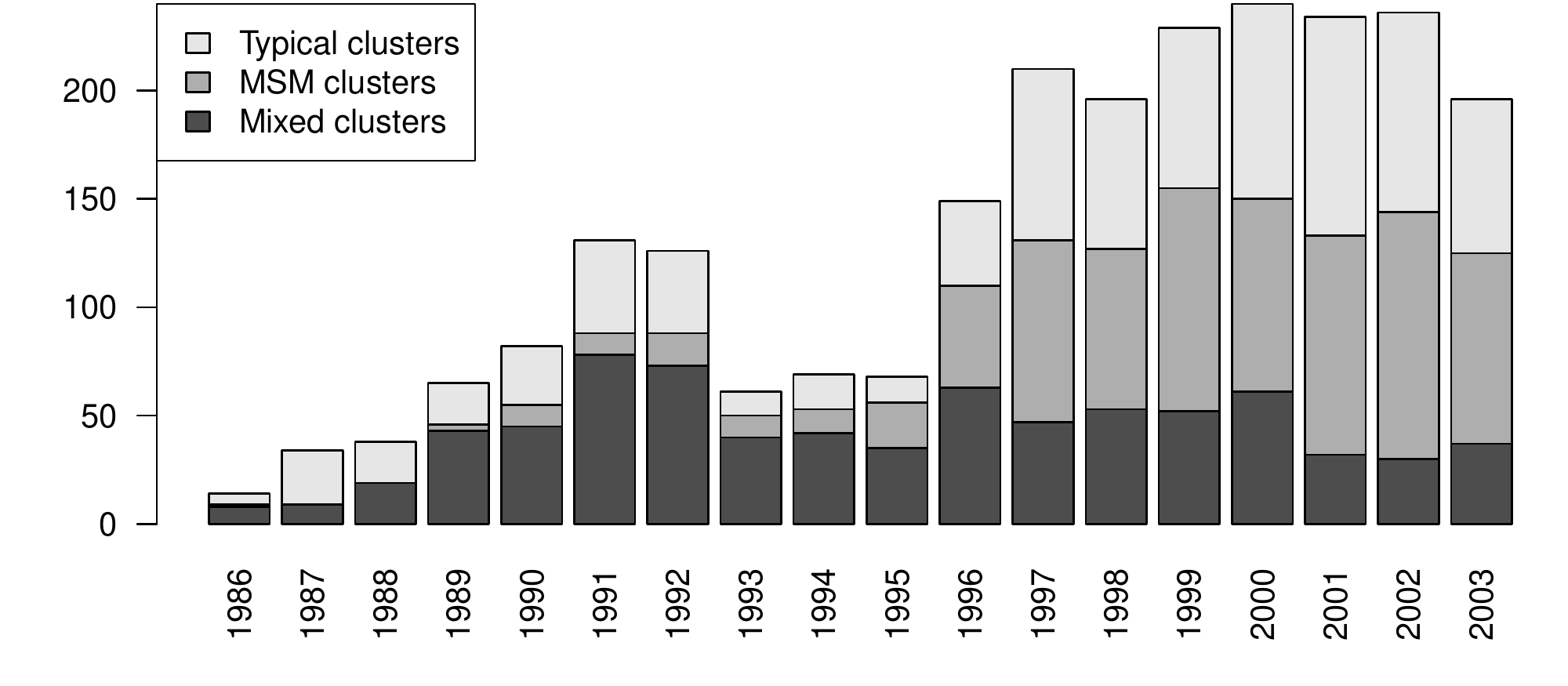}
  \caption{Yearly distribution of the three groups of clusters}
  \label{fig:groups:year}
\end{figure}

This analysis shows that the two groups made of atypical clusters are far from
each other compared to their internal distances. This is confirmed by the
detection date analysis displayed on Figure \ref{fig:groups:year}. It appears
that the epidemic in the giant component has two separated components. One
mostly male homosexual component tends to dominate the recent cases 
(note that even typical clusters contain at least 57 \% of MSM), while a mixed
component with a large percentage of female patients was dominating the early
epidemic, but tends to diminish recently. It should also be noted that this
mix component is dominated by the growth of the homosexual
component, but seems to decay only slightly in absolute terms. In other words,
the reduction should be seen as an inability to control the growth homosexual
epidemic rather than as a success in eradicating the heterosexual epidemic. 

\section{Conclusion}
The proposed visual mining method for graphs has been shown to provide
valuable insights on the epidemic network. It is based on links between
modularity and visualization and leverages recent computationally efficient
modularity maximizing methods. Future works include the integration of the
proposed methods in graph mining tools such as \cite{ICWSM09154} and its
validation on other aspects of epidemic networks analysis.

\end{document}